\RequirePackage{ifpdf}
\documentclass[nohyper,12pt,letterpaper]{JHEP3}
\usepackage{epsfig}
\usepackage[latin1]{inputenc}
\usepackage{bbm,amsfonts}
\usepackage{graphicx}
\usepackage{amssymb,amsmath}
\usepackage{fancybox,framed}
\usepackage{dsfont}
\usepackage{mathtools}
\usepackage{braket}
\usepackage{cite}
\usepackage{caption}
\usepackage{subcaption}

\author{Marco S. Bianchi$^{\bf a}$
  and Matias Leoni$^{\bf b,c}$\\\\
 $^{\bf a}$ Centre for Research in String Theory,
School of Physics and Astronomy\\
Queen Mary University of London,
Mile End Road, London E1 4NS, UK \\
  $^{\bf b}$  Physics Department, FCEyN-UBA \& IFIBA-CONICET\
Ciudad Universitaria, Pabell\'on I, 1428, Buenos Aires, Argentina \\
  $^{\bf c}$  Instituto de F\'isica de La Plata, CONICET, UNLP\
C.C. 67, 1900 La Plata, Argentina \\
  \qquad\\
  E-mail: \email{ m.s.bianchi@qmul.ac.uk, 
  leoni@df.uba.ar}
}

\abstract{We consider a light-like Wilson loop in ${\cal N}=4$ SYM evaluated on a regular $n$-polygon contour. Sending the number of edges to infinity the polygon approximates a circle and the expectation value of the light-like WL is expected to tend to the localization result for the circular one. We show this explicitly at one loop, providing a prescription to deal with the divergences of the light-like WL and the large $n$ limit.
Taking this limit entails evaluating certain sums of dilogarithms which, for a regular polygon, evaluate to the same constant independently of $n$. We show that this occurs thanks to underpinning dilogarithm identities, related to the so-called ``polylogarithm ladders'', which appear in rather different contexts of physics and mathematics and enable us to perform the large $n$ limit analytically.
}

\preprint{November 2014\\QMUL-PH-14-26}

\title{Dilogarithm ladders from Wilson loops}


\csname @addtoreset\endcsname{equation}{section}

\def\bseq{\begin{subequation}}  
\def\eseq{\end{subequation}}
\def\bsea{\begin{subeqnarray}}  
\def\esea{\end{subeqnarray}}

\newcommand{\beq}{\begin{equation}}
\newcommand{\bea}{\begin{eqnarray}}
\newcommand{\eea}{\end{eqnarray}}
\newcommand{\eeq}{\end{equation}}

\def\beq{\begin{equation}}
\def\eeq{\end{equation}}
\def\bea{\begin{eqnarray}}
\def\eea{\end{eqnarray}}

\begin{document}

\allowdisplaybreaks

\section{Introduction}

In this note we consider a limit on light-like Wilson loops on a regular polygonal contour in ${\cal N}=4$ SYM.
By regular we mean that all the light-like sides of the polygon have equal euclidean norm.
Then, sending the number of edges to infinity, we expect the contour to approximate a smooth circle. Consequently, the expectation value of the light-like Wilson loop is supposed to reproduce that of the $\tfrac12$-BPS circular one, whose exact expression is known from a matrix model computation \cite{Erickson:2000af} as a result of localization \cite{Pestun:2007rz}.

Light-like Wilson loops in ${\cal N}=4$ SYM are dual to MHV scattering amplitudes \cite{Drummond:2007cf,Drummond:2007au,Drummond:2008aq,Brandhuber:2007yx}.
The fact that the limit we are taking on the light-like Wilson loop is known, implies in  turn that the same limit should also hold for MHV scattering amplitudes, with special kinematics.
This could be in principle used as a check on potential expressions for MHV scattering amplitudes of ${\cal N}=4$ SYM for any number of external particles, though the limit could be hard to perform analytically in practice.
Nevertheless, a similar argument was used in \cite{Alday:2007he} (approximating a rectangular Wilson loop by a sequence of light-like segments) to find an inconsistency of the BDS proposal \cite{Bern:2005iz} for multileg amplitudes at strong coupling \cite{Alday:2007hr}.

Despite the simplicity of this idea we are not aware of any direct computation in the literature where such kind of limit has been explicitly checked perturbatively.
Here we provide an explicit computation of the expectation value of a ${\cal N}=4$ SYM light-like Wilson loop approximating the circular one, at one loop in perturbation theory.

Light-like Wilson loops suffer from ultraviolet divergences, which are dual to infrared singularities of scattering amplitudes. At one loop, these arise from gluon exchanges between adjacent edges. Therefore one has to introduce a regularization to deal with them. In particular we use dimensional regularization.
On the other hand the $\tfrac12$-BPS Wilson loop evaluated on a smooth contour, such as a circle, is finite \cite{Erickson:2000af}.
In interpolating between the two the limit of infinite number of edges is taken. We give a prescription to deal with this limit in such a way that the result is finite, as expected.
This effectively removes any contribution from divergent diagrams, leaving a sum over finite ones, where we can set the regularization parameter to zero.
Remarkably, once this is done, and for the particular regular contour we have chosen, the result of the sum over these one-loop contributions turns out to be constant, i.e. independent of $n$.
Since the integrals involved in the computation are expressible in terms of dilogarithms, this implies in turn identities for the sum of dilogarithms evaluated at particular values of their arguments.
When these are powers of the same algebraic number, such identities are called dilogarithm ladders and there exists a vast literature on these (see \cite{Lewin,Lewin2} for a review, or \cite{Cohen} and \cite{Broadhurst} for more recent developments).
This is not precisely the kind of identities we encounter, though they should be related by use of the Abel identity.  
Nevertheless, other identities relating combinations of dilogarithms were studied in the past, emerging from various problems of mathematical physics, such as in the context of Heisenberg spin chains \cite{KR1}, two-dimensional integrable and lattice models \cite{BR,R,Zm} and CFT's \cite{Nahm:1992sx,DKKMM} whose central charge can be expressed as a sum of dilogarithms evaluated at particular algebraic numbers\footnote{For a more comprehensive list of references and applications of dilogarithms in physics and mathematics we refer the reader to \cite{Kirillov:1994en} and the references therein.}. 
We find that these dilogarithm identities, which have been proved \cite{Kir4,Kir7}, include those we face in our computation as particular subcases. 
Thanks to these results we can analytically prove that the limit on the regular polygon Wilson loops converges to the one-loop expansion of the localization expression for the circular one.

At one loop one can straightforwardly compute the expectation value of the circular Wilson loop by a Feynman diagram computation. Insisting on using dimensional regularization we can extend the comparison of the light-like and circular Wilson loops to finite values of the regularization parameter $\epsilon$.
In this case we still verify that the finite part of the light-like Wilson loop converges to the circular one in the large $n$ limit, for any value of $\epsilon$, although the two results do not coincide at finite $n$, in contrast with the $\epsilon\rightarrow 0$ limit.

\section{Contour parametrization}

In this section we derive a parametrization of the polygonal contour with $n$ edges, on which we want to evaluate the light-like Wilson loop.
We restrict our analysis to regular light-like polygons ${\cal P}= \bigcup_{i=1}^n x_{i,i+1}$ with an even number of edges $n$.
We parametrize their vertices as follows
\begin{align}\label{eq:parametrization}
x_{2k+1} & = \left( 2 \sin \frac{\pi}{n}, \sin \frac{2\pi (2k+1)}{n} , \cos \frac{2\pi (2k+1)}{n} \right)
\nonumber\\
x_{2k} & = \left( 0, \sin \frac{4\pi k}{n} , \cos \frac{4\pi k}{n} \right) \qquad ,\qquad k = 0,1, \dots n/2-1
\end{align}
\begin{figure}
\centering
\begin{subfigure}{.5\linewidth}
  \centering
  \includegraphics[width=0.6\linewidth]{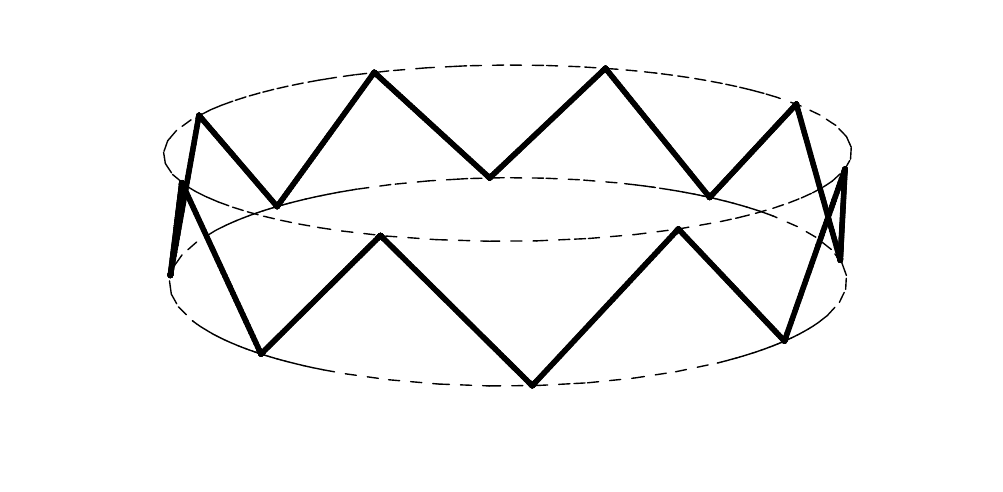}
  \caption{$n=16$}
\end{subfigure}%
\begin{subfigure}{.5\linewidth}
  \centering
  \includegraphics[width=0.6\linewidth]{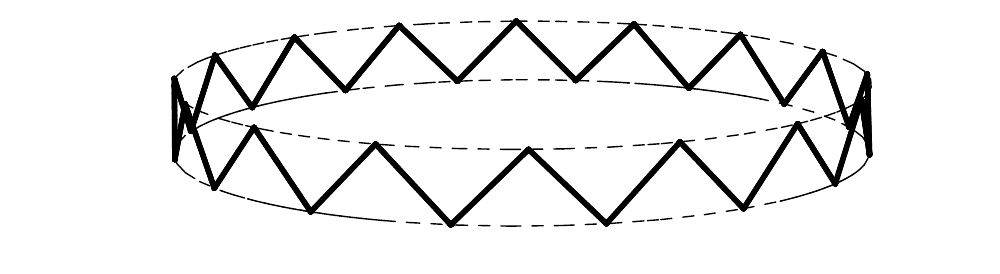}
  \caption{$n=32$}
\end{subfigure}
\begin{subfigure}{.5\linewidth}
  \centering
  \includegraphics[width=0.6\linewidth]{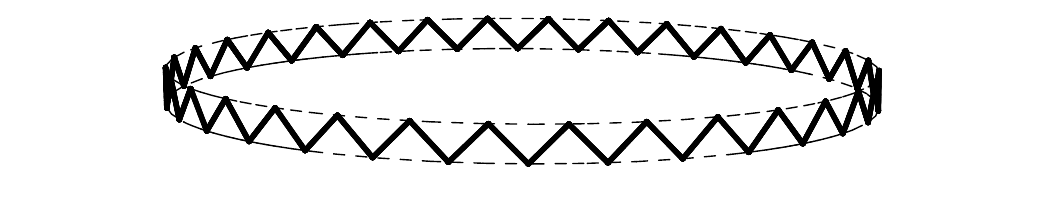}
  \caption{$n=64$}
\end{subfigure}
\caption{Examples of the contour for $n=16$, $32$ and $64$. The polygon is formed by connecting with light-like lines two sets of points lying on a circle at $t=0$ and at $t=2\sin\tfrac{\pi}{n}$ respectively ($t$ is the vertical axis in the figures). In the limit $n\to\infty$ the contour becomes a space-like circle of radius $1$. }
\label{fig:polygon}
\end{figure}
This means that points with odd and even labels lie on two circles of equal radii and on two parallel space-like planes separated by a distance $t = 2 \sin \frac{\pi}{n}$ in the time direction.
The overall radii of the circles do not play any role thanks to conformal invariance and are set to unity (see Figure \ref{fig:polygon}). 

Given this parametrization, the relevant invariants on which the Wilson loop can depend on can be separated into two categories: odd-to-odd and odd-to-even distances (even-to-even are equal to odd-to-odd by symmetry). They are evaluated from \eqref{eq:parametrization}
\begin{align}
x^2_{2k,2l} &= 4 \sin^2 \frac{2\pi (k-l)}{n}
\nonumber\\
x^2_{2k,2l+1} &= 4 \sin^2 \frac{\pi (2k-2l-1)}{n} - 4\sin^2 \frac{\pi}{n} 
\end{align}
For a polygon with $n$ edges there are $n/4$ odd-to-odd and $(n/4-1)$ odd-to-even independent distances if $n=0$ $mod$ 4 and $(n-2)/4$ odd-to-odd and odd-to-even independent distances if $n=2$ $mod$ 4, as reviewed in Table \ref{table:invariants}.

\begin{table}[h]
\begin{center}
    \begin{tabular}{| l || l | l || l | l |}
    \hline
    n & $x_{2k,2l}$ & $x_{2k,2l+1}$ & $\uparrow \uparrow + \downarrow \downarrow$ & $\uparrow \downarrow + \downarrow \uparrow$  \\ \hline
    4 & 1 & 0 & 2 & 4 \\ \hline
    6 & 1 & 1 & 6 & 9 \\ \hline
    8 & 2 & 1 & 12 & 16 \\ \hline
    \dots & \dots & \dots & \dots & \dots \\ \hline
    $4m$ & $m$ & $m-1$ & $2m(2m-1)$ & $4m^2$  \\ \hline
    $4m+2$ & $m$ & $m$ & $2m(2m+1)$ & $(2m+1)^2$ \\
    \hline
    \end{tabular}
    \caption{For different number of edges we report the number of independent invariants (central columns) and the number of diagrams with gluon exchanges between edges separated by an odd (segments pointing both to the future or to the past) or even (segments pointing in opposite directions in time) number of sides.}
    \label{table:invariants}
    \end{center}
\end{table}

\section{One-loop integrals}

\begin{figure}
\centering
\includegraphics[width=0.5\textwidth]{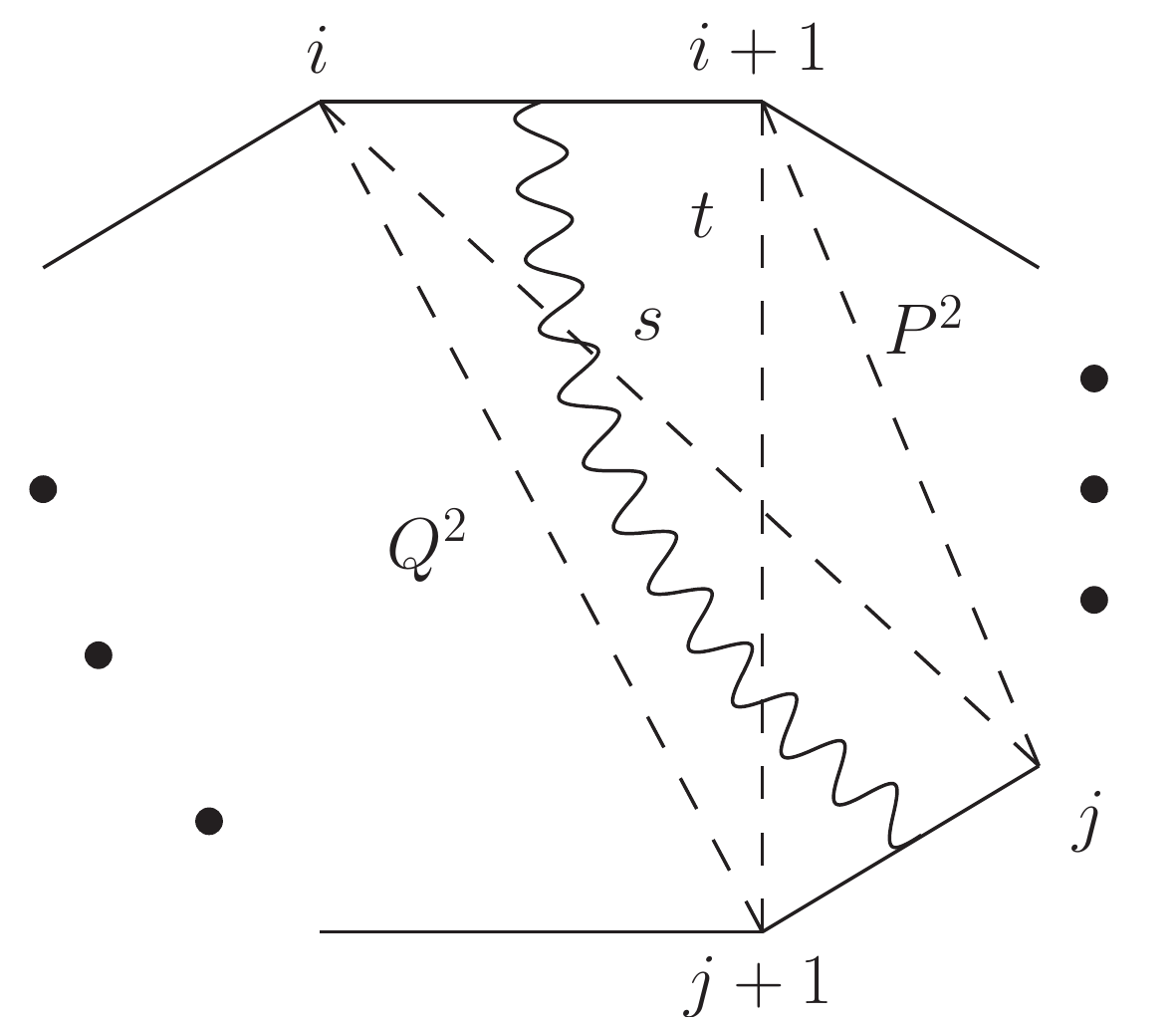}
\caption{One-loop contribution to the Wilson loop.}
\label{fig:WL1loop}
\end{figure}

In this section we evaluate the expectation value of the light-like Wilson loop on the polygon ${\cal P}$ at one loop in perturbation theory.
We use the position space gluon propagator in dimensional regularization
\begin{equation}
\langle A_{\mu}(x) A_{\nu}(y) \rangle = -\frac{\Gamma(1-\epsilon)}{4\pi^{2-\epsilon}}\, \frac{\eta_{\mu\nu}}{\left[-(x-y)^2\right]^{1-\epsilon}}
\end{equation}
where $\eta_{\mu\nu}$ is the Minkowski metric.
An overall factor $(i g)^2 N = \lambda$ will be understood in the following.
There is only one kind of integral to be considered at one loop \cite{Brandhuber:2007yx}
\begin{equation}\label{eq:integrand}
I^{(1)}_{i,j} \equiv \frac{\Gamma(1-\epsilon)}{8\pi^{2-\epsilon}}\! \int_0^1\!\! d\tau_i\, d\tau_j\, \frac{P^2+Q^2-s-t}{\left[-\left(P^2+\tau_i(s-P^2)+\tau_j(t-P^2)+(P^2+Q^2-s-t)\tau_i\tau_j\right)\right]^{1-\epsilon}}
\end{equation}
which is conveniently expressed in terms of the distances $s\equiv x_{i,j}^2$, $t \equiv x_{i+1,j+1}^2$, $P^2\equiv x_{i+1,j}^2$ and $Q^2\equiv x_{i,j+1}^2$, as in Figure \ref{fig:WL1loop}.
Its limit where $s,t,P^2\rightarrow 0$, namely whenever the gluon is exchanged between two adjacent edges, yields divergent contributions which can be evaluated in dimensional regularization
\begin{equation}\label{eq:div}
I^{(1)}_{i,i+1} = -\frac12\, \frac{\Gamma(1-\epsilon)}{4\pi^{2-\epsilon}}\, \frac{(-Q^2)^\epsilon}{\epsilon^2}
\end{equation}
Whenever a gluon is exchanged between two edges separated by only one light-like side, we have, e.g., $P^2=0$. The corresponding contribution can be smoothly obtained as a limit of the integral \eqref{eq:integral}. For nonvanishing invariants this integral is finite and can be evaluated at $\epsilon=0$ 
\begin{equation}\label{eq:integral}
I^{(1)}_{i,j} = \frac{1}{8\pi^2}\, \left[ -{\rm Li}_2 (1-as) - {\rm Li}_2 (1-at) + {\rm Li}_2 (1-aP^2) + {\rm Li}_2 (1-aQ^2) \right]
\end{equation}
where
\begin{equation}
a \equiv \frac{P^2 + Q^2 - s - t}{P^2 Q^2 - s t}
\end{equation}
Using regular polygons yields further simplifications.
First, thanks their symmetries we have that $s=t$ in each exchange.
Second, the integral only depends on the integer number $\Delta i$, which is the difference between the labels of the two edges, between which the gluon is exchanged.
We will call it $\Delta i$, hereafter.
At one loop, there is a distinction between gluon exchanges connecting edges pointing in the same time direction or pointing in the opposite.
Let us analyze the latter case first: we have that $P^2 Q^2 = s t$, meaning that $a$ diverges.
The limit $a\to\infty$ cannot be taken from the result \eqref{eq:integral}.
Rather, it is more convenient to implement this condition on \eqref{eq:integrand} directly. When this is done the integrand factorizes and can be integrated straightforwardly
\begin{equation}\label{eq:even-to-odd}
I^{(1)}_{2k,2l+1} = I^{(1)}_{2k+1,2l} = -\frac12 \frac{\Gamma(1-\epsilon)}{4\pi^{2-\epsilon}} \frac{(-P^2)^{-\epsilon}\, \left[ \left(\frac{P^2}{s}\right)^\epsilon -1 \right]^2}{\epsilon^2}
\end{equation}
Even though the result is finite, we evaluate it in dimensional regularization for future convenience.

Finally, the case where the gluon is exchanged between two edges pointing in the same direction is completely regular and we can use \eqref{eq:integral} directly.

As a whole we have
\begin{equation}
I^{(1)}(\Delta i) = \left\{
\begin{array}{ll}
-\frac12\, \frac{\Gamma(1-\epsilon)}{4\pi^{2-\epsilon}\, \epsilon^2}\, \left(4 \sin^2 \frac{2\pi}{n} \right)^{\epsilon} & \Delta i = 1 \\
\\
-\frac12 \frac{\Gamma(1-\epsilon)}{4\pi^{2-\epsilon}\, \epsilon^2} \left(4\sin^2 \frac{\pi (\Delta i-1)}{n} \right)^{-\epsilon}\, \left[ \left(\frac{\sin^2 \frac{\pi (\Delta i-1)}{n}}{\sin^2 \frac{\pi \Delta i}{n}-\sin^2 \frac{\pi}{n}}\right)^\epsilon -1 \right]^2  & \Delta i \, odd \\
\\
\frac{1}{8\pi^2}\, \left[  {\rm Li}_2 \left( \frac{\cos^2 \frac{\pi (\Delta i + 1)}{n}}{\cos^2 \frac{\pi}{n}} \right) + {\rm Li}_2 \left( \frac{\cos^2 \frac{\pi (\Delta i - 1)}{n}}{\cos^2 \frac{\pi}{n}} \right) - 2 {\rm Li}_2 \left( 1-\frac{\sin^2 \frac{\pi \Delta i}{n}}{\cos^2 \frac{\pi}{n}} \right) \right] & \Delta i \, even \\
\end{array}
\right.
\end{equation}

\section{Dilogarithm identities and circular limit}

With a little combinatorics on the polygons, we can simplify the final expression for the one-loop correction of the Wilson loop expectation value.
Gluons can be exchanged between edges separated by $\Delta i = 1,2, \dots n/2$. Even and odd separations correspond to different cases and are treated separately.
For each separation there are $n$ different gluon exchanges, apart from the extremal case $\Delta i = n/2$, where there are only $n/2$. Thanks to the symmetry of the polygons, all contributions within edges at the same separation $\Delta i$ are equal to each other.
This can be used to reduce the double sums over the indices of the edges of gluon exchanges, implicit in the total contribution, to single sums over the separations.
Then the combinatorics vary according to whether the total number of edges $n=0$ $mod$ 4 or $n=2$ $mod$ 4, as reviewed in the Table \ref{table:invariants}.

The expression for the one-loop Wilson loop expectation value reads in all cases
\begin{align}\label{eq:total}
& \langle W \rangle^{(1)} =  
- n \frac12\, \frac{\Gamma(1-\epsilon)}{4\pi^{2-\epsilon}\, \epsilon^2}\, \left(4 \sin^2 \frac{2\pi}{n} \right)^{\epsilon} + 
\nonumber\\&
- \frac{n}{2}\, \sum_{k=1}^{n/2-2} \frac12 \frac{\Gamma(1-\epsilon)}{4\pi^{2-\epsilon}\, \epsilon^2} \left(4\sin^2 \frac{2\pi k}{n} \right)^{-\epsilon}\, \left[ \left(\frac{\sin^2 \frac{2\pi k}{n}}{\sin^2 \frac{\pi (2k+1)}{n}-\sin^2 \frac{\pi}{n}}\right)^\epsilon -1 \right]^2 +
\nonumber\\&
+ \frac{n}{2}\, \sum_{k=1}^{n/2-1} 
\frac{1}{8\pi^2}\, \left[ {\rm Li}_2 \left( \frac{\cos^2 \frac{\pi (2k + 1)}{n}}{\cos^2 \frac{\pi}{n}} \right) + {\rm Li}_2 \left( \frac{\cos^2 \frac{\pi (2k - 1)}{n}}{\cos^2 \frac{\pi}{n}} \right) - 2 {\rm Li}_2 \left( 1-\frac{\sin^2 \frac{2\pi k}{n}}{\cos^2 \frac{\pi}{n}} \right) \right]
\end{align}
In the first line the $n$ divergent contributions from exchanges between adjacent edges appear.
The second line represents the remaining $n/2(n/2-2)$ terms from even-to-odd exchanges.
They sum up for a total of $(n/2)^2$ terms, as in the last column of Table \ref{table:invariants}.
Finally the third line is given by the $n/2(n/2-1)$ contributions from odd-to-odd and even-to-even contributions, as in the fourth column of Table \ref{table:invariants}. 

Next we take the $n\to\infty$ limit of such an expression. In this regime, we expect the light-like polygonal contour to approximate a circle.
The supersymmetric Wilson loop on a circle has been given an exact result through localization on a four-sphere, yielding a Gaussian matrix model.
Such a Wilson loop operator features a coupling to the scalar fields of ${\cal N}=4$ SYM. On the other hand, for a light-like contour such terms are dropped and one recovers the standard Wilson loop. Even with this difference in the couplings, we expect the expectation value of the light-like polygonal Wilson loop to tend to that of the circular one in the large $n$ limit.

Since the circular Wilson loop is finite, we expect the ultraviolet divergences from the cusps of the light-like Wilson loop to drop out in such a regime.
In order to ensure this, we take this limit on the dimensionally regularized result \eqref{eq:div}.
In the $n\to\infty$ limit the invariants vanish and the relevant integral becomes scaleless and is thus discarded in dimensional regularization. This result is still to be summed on order $n$ terms with $n$ tending to infinity, meaning that choosing a correct prescription amounts to an order of limits problem. We take the $n\rightarrow\infty$ first, then in the spirit of dimensional regularization we analytically continue the result from the region in $\epsilon$ space where the sum converges. Eventually we take the $\epsilon\to 0$ limit, which vanishes.

Considering the even-to-odd contribution \eqref{eq:even-to-odd} we can perform the same reasoning, namely taking the limit on the invariants before sending $\epsilon\to 0$, which ensures that these terms also go to 0 in the large $n$ limit.

Finally the most interesting contribution comes from the even-to-even terms.
Indeed this piece evaluates to a constant for any value of $n$, namely
\begin{equation}\label{eq:identity}
\boxed{\frac{n}{2}\, \sum_{k=1}^{n/2-1} 
\left[ {\rm Li}_2 \left( \frac{\cos^2 \frac{\pi (2k + 1)}{n}}{\cos^2 \frac{\pi}{n}} \right) + {\rm Li}_2 \left( \frac{\cos^2 \frac{\pi (2k - 1)}{n}}{\cos^2 \frac{\pi}{n}} \right) - 2 {\rm Li}_2 \left( 1-\frac{\sin^2 \frac{2\pi k}{n}}{\cos^2 \frac{\pi}{n}} \right) \right] = \pi^2}
\end{equation}
Surprisingly, even if one would have expected an $n$-dependent expression which should evaluate to $\pi^2$ in the limit of large $n$, a numerical inspection shows that the left hand side of (\ref{eq:identity}) evaluates to $\pi^2$ for any value of $n$. For the moment this is just a case by case empirical observation that we extrapolate to be valid for $n\rightarrow\infty$ (we shall later prove this is actually correct by recasting the result in terms of know identities).
In conclusion, performing the large $n$ limit is trivial and, taking into account the $1/(8\pi^2)$ factor in \eqref{eq:total}, the expectation value of the Wilson loop gives $1/8$.
This coincides with the first order contribution to the circular Wilson loop at weak coupling
\begin{equation}
\langle W_{\bigcirc} \rangle = 1 + \frac{\lambda}{8} + \dots
\end{equation}

Besides triggering the limit of the light-like Wilson loop to the circular one, the identity \eqref{eq:identity} is very interesting by itself, since it relates combinations of dilogarithms at particular values of their arguments.
Such relations are of mathematical interest and there exists a vast literature on them. When they relate polylogarithms of the same weight evaluated at powers of some algebraic number $\phi$, they are known as polylogarithm ladders.
Such identities are usually written more compactly in terms of the Rogers function ${\rm L}$ which is defined to be
\begin{equation}
{\rm L}(x) \equiv {\rm Li}_2(x) + \frac12\, \log x\, \log(1-x)
\end{equation}
for $0\leq x \leq 1$ and analytically continued to the other regimes of $x$ through the reflection and inversion identities
\begin{align}
&{\rm L}(x) = \frac{\pi^2}{3} - {\rm L}\left(\frac{1}{x}\right) \qquad x>1 \nonumber\\&
{\rm L}(x) = {\rm L}\left(\frac{1}{1-x}\right)-\frac{\pi^2}{6} \qquad x<0
\end{align}
It satisfies the Abel identity
\begin{equation}
{\rm L}(x) + {\rm L}(y) = {\rm L}(xy) + {\rm L}\left(\frac{x(1-y)}{1-xy}\right) + {\rm L}\left(\frac{y(1-x)}{1-xy}\right)
\end{equation}
for $0<x,y<1$.
As an example of an early dilogarithm ladder we quote one of Watson's identities \cite{W}
\begin{equation}\label{eq:Watson}
2{\rm L}(\beta) + {\rm L}(\beta^2) = \frac{10}{7}\, L(1)
\end{equation}
where $\beta=\frac12\, \sec{\frac{\pi}{7}}$ is one of the roots of the equation 
\begin{equation}
x^3 - 2x^2 -x +1 = 0
\end{equation}

We can now inspect formula \eqref{eq:identity} in order to see if it can be related to known identities in such a way to analytically prove the agreement with the circular Wilson loop.
It is convenient to analyze the $n=2\, (mod\, 4)$ and $n=0\, (mod\, 4)$ cases separately\footnote{In the following steps we also find convenient to send $n\rightarrow 2n$.}.
From the former case we can massage \eqref{eq:identity} (rearranging the sum) to obtain the equivalent identity
\begin{equation}\label{eq:identityRogers2}
\sum_{k=1}^{n/2-1/2}\, {\rm L}\left( \frac{\sin^2 \frac{k}{n}\pi}{\cos^2 \frac{\pi}{2n}} \right) - \frac{\pi^2}{24 n} \left( n^2 + 3 \right) = 0 \qquad n=1\, (mod\, 2)
\end{equation}
In this form we can compare it to the result by Kirillov \cite{Kir7} (eq. (1.16) of \cite{Kirillov:1994en}, properly rewritten)
\begin{align}
& \sum_{k=1}^{n/2-1/2}\, {\rm L}\left( \frac{\sin^2 \frac{(j+1)\pi}{n}}{\sin^2 \frac{k(j+1)\pi}{n}} \right) - \frac{\pi^2}{6 n} \left( (3j+1)(n-2) - 3j^2 -1 + n \right) = 0 \\& 
n=1\, (mod\, 2) \qquad 0 \leq j \leq \frac{n-3}{2} \qquad g.c.d (n,j+1) = 1 \nonumber
\end{align}
We specialize to the case $j= \frac{n-3}{2}$ and, using the inversion functional equation of the Rogers function, 
and after some further rearrangements of the sum, it can be brought to the form \eqref{eq:identityRogers2}, thus proving their equivalence. 

In the $n=0$ $(mod\, 4)$ case, using again functional identities of the Rogers function and manipulations of the sum, we can derive from \eqref{eq:identity} the equivalent identity
\begin{equation}
\sum_{k=1}^{\frac{n-2}{2}}\, {\rm L}\left( \frac{\cos^2 \frac{k}{n}\pi}{\cos^2 \frac{\pi}{n}} \right)
- \frac12\, {\rm L}\left( \cos^2 \frac{\pi}{n} \right) + \frac{\pi^2}{24 n} \left( n^2 - 4n + 12 \right) = 0 \qquad n=0\, (mod\, 2)
\end{equation}
This can be shown to coincide with a special case of Kirillov's result \cite{Kir7} (eq. (1.28) of \cite{Kirillov:1994en})
\begin{equation}
\sum_{k=1}^{n-2}\, {\rm L}\left( \frac{\sin^2 \frac{(j+1)\pi}{n}}{\sin^2 \frac{(k+1)(j+1)}{n}\pi} \right) = \frac{\pi^2}{6}\, \left( \frac{3n-6}{n} - \frac{6 j (j+2)}{n} + 6 \mathbb{Z}_+ \right)
\end{equation}
setting $j=n/2-2$ and with some rewriting. In this case the positive integer number is found to be $j$.

From these formulae it is possible to derive some classic ladder identities, by using the Abel identity and the algebraic equations satisfied by the algebraic numbers appearing in them.
For instance Watson's identity \eqref{eq:Watson} can be obtained from \eqref{eq:identityRogers2}, taking $n=7$ with some manipulations involving the Abel identity.

In other words, using dilogarithm identities allows us to prove that the limit of the light-like Wilson loop expectation value coincides with that of the circular one at one loop.
It is amusing that in our setting this limit is driven by nontrivial dilogarithm identities which appeared in rather disparate contexts in theoretical physics.

One could also think the phenomena we found the other way around. We have chosen a contour which is parametrized by trigonometric functions evaluated in rational portions of $\pi$ which are algebraic numbers. Since the finite pieces of the result for the Wilson loop are formed by a uniform transcendentality combination of polylogarithmic functions, we have somehow ``generated'' a non trivial set of polylogarithm identities of algebraic numbers. It would be interesting to understand if there is an underlying explanation of such a surprising result and if it would be possible to generalize it to other contours and higher loops leading to identities about which much less is known.

\section{Higher order in dimensional regularization parameter}

The one-loop expectation value of both the circular and light-like Wilson loops can be derived at finite $\epsilon$.

In the former case we have
\begin{align}
\langle W_{\bigcirc} \rangle^{(1)} = \frac{2^{2 \epsilon -3} \pi ^{\epsilon -\frac{1}{2}} \Gamma (1-\epsilon ) \Gamma \left(\epsilon +\frac{1}{2}\right)}{\Gamma (\epsilon +1)}
\end{align}
this is the result of a straightforward Feynman diagram computation which consistently coincides with the localization result at $\epsilon=0$.

For the light-like Wilson loop we can take the $n=2$ $(mod\, 4)$ case for simplicity and compute the relevant integral \eqref{eq:integrand} to all order in $\epsilon$.
This yields
\begin{align}
- \frac12\, \frac{\Gamma (1-\epsilon )}{4\pi^{2-\epsilon}}&\, \frac{a}{\epsilon  (\epsilon +1)} \left((P^2)^{\epsilon +1} \, _2F_1\left(1,\epsilon +1;\epsilon +2; a\, P^2\right) + \right. \nonumber\\& \left.
 + (Q^2)^{\epsilon +1} \, _2F_1\left(1,\epsilon +1;\epsilon +2; a\, Q^2\right) - 2 s^{\epsilon +1} \, _2F_1\left(1,\epsilon +1;\epsilon +2;a\, s\right)\right)
\end{align}
except for the case where $s=P^2$ (or $s=Q^2$), which appears on the regular polygon for $k=\frac{n-2}{4}$ and $k=\frac{n+2}{4}$ respectively, where the integral becomes
\begin{align}
\frac12\, \frac{\Gamma (1-\epsilon )}{4\pi^{2-\epsilon}}\, (Q^2-s) s^{\epsilon -1} \, _3F_2\left(1,1,1-\epsilon ;2,2;1-\frac{Q^2}{s}\right)
\end{align}
Performing the relevant sum \eqref{eq:total} we can compute numerically the one-loop expectation value as a function of $n$ and $\epsilon$. 
At fixed finite $\epsilon$ the sums are not constant any longer, nevertheless for large values of $n$ one can verify that the light-like Wilson loop approximates the circular Wilson loop expectation value at any value of $\epsilon$.

\begin{figure}
\centering
\includegraphics[width=0.6\textwidth]{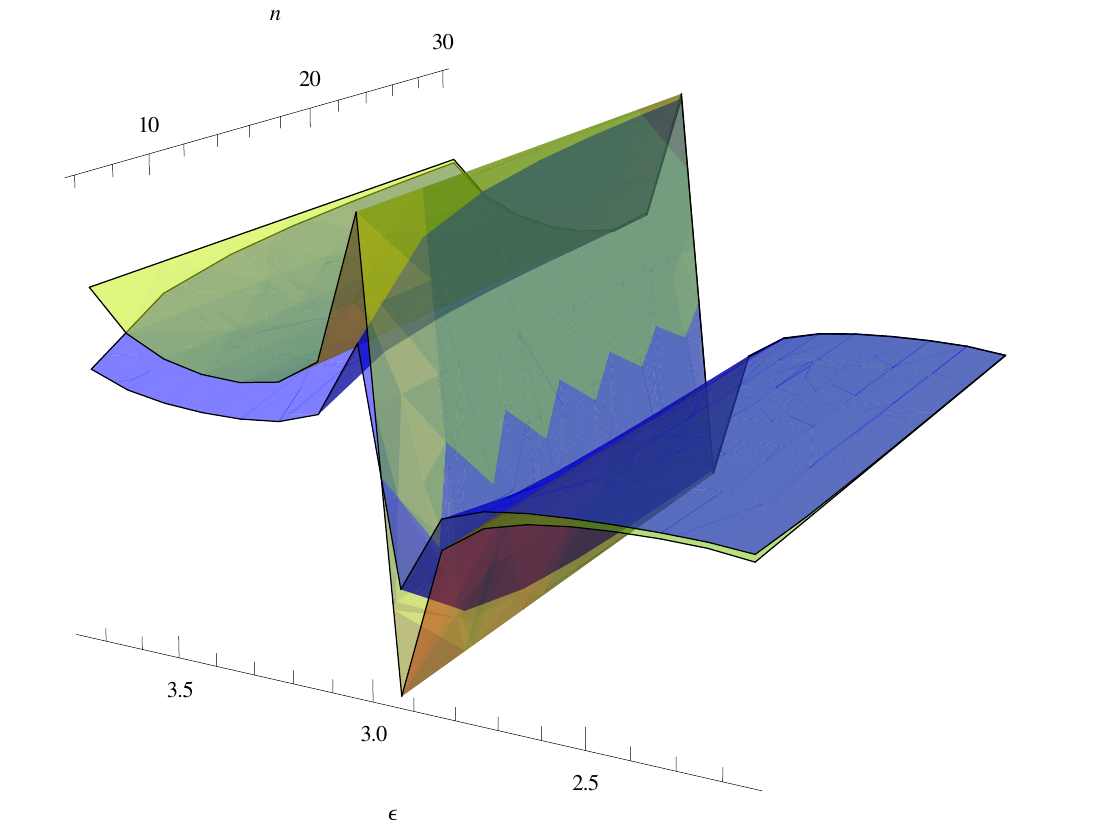}
\caption{Comparison between the one-loop expectation values of the light-like (blue) and circular (yellow) Wilson loops, as functions of the number of sides $n$ and the regularization parameter $\epsilon$. For the polygonal Wilson loop the surface is an extrapolation between the discrete points. At large $n$ the two surfaces coincide.}
\label{fig:comparison}
\end{figure}

As an example Figure \ref{fig:comparison} shows a plot of the relevant part of the light-like Wilson loop expectation value (blue) against the circular one (yellow) as a function of $n$ and $\epsilon$. For large $n$ the two surfaces coincide.

\section*{Acknowledgements}

We thank Lorenzo Bianchi, Gast\'on Giribet and Gabriele Travaglini for very useful discussions. Part of this work was performed at the 2014 Spring School on Superstring Theory and Related Topics held at ICTP. We want to thank the organizers for providing the stimulating environment. This work was supported in part by the Science and Technology Facilities Council Consolidated Grant ST/L000415/1 \emph{String theory, gauge theory \& duality}.

\vfill
\newpage

\bibliographystyle{JHEP}

\bibliography{biblio}

\end{document}